**Title**: A hidden Markov modeling approach combining objective measure of activity and subjective measure of self-reported sleep to estimate the sleep-wake cycle

Semhar B Ogbagaber[1*], Yifan Cui[2], Kaigang Li[3], and Ronald J Iannotti[4], and Paul S Albert[5]

*Corresponding author

[1] Bristol Myers Squibb, Lawrenceville, NJ, USA
[2] Center for Data Science, Zhejiang University, China
[3] Colorado School of Public Health, Aurora, CO, USA
[4] The CDM Group, Bethesda, MD, USA
[5] Division of Cancer Epidemiology and Genetics, National Cancer Institute, Bethesda, MD, USA

**Abstract**

Characterizing the sleep-wake cycle in adolescents is an important prerequisite to better understand the association of abnormal sleep patterns with subsequent clinical and behavioral outcomes. The aim of this research was to develop hidden Markov models (HMM) that incorporate both objective (actigraphy) and subjective (sleep log) measures to estimate the sleep-wake cycle using data from the NEXT longitudinal study, a large population-based cohort study. The model was estimated with a negative binomial distribution for the activity counts (1-minute epochs) to account for overdispersion relative to a Poisson process. Furthermore, self-reported measures were dichotomized (for each one-minute interval) and subject to misclassification. We assumed that the unobserved sleep-wake cycle follows a two-state Markov chain with transitional probabilities varying according to a circadian rhythm. Maximum-likelihood estimation using a backward-forward algorithm was applied to fit the longitudinal data on a subject by subject basis. The algorithm was used to reconstruct the sleep-wake cycle from sequences of self-reported sleep and activity data. Furthermore, we conduct simulations to examine the properties of this approach under different observational patterns including both complete and partially observed measurements on each individual.

# 1. Introduction

Requirements for daily periods of sleep decrease with age from 13 hours/day for infants and 9 hours/day for adolescents to 7 to 8 hours/day of sleep recommended for adults [32]. The issue of estimating the sleep-wake cycle has been of increased interest as researchers have identified associations between abnormal sleeping patterns and subsequent clinical and behavioral outcomes in both adults [24, 34] and children [10, 14]. Characterizing the circadian cycle could be beneficial in the development of interventions for people suffering with various forms of sleep disorders.

Actigraphy and self-reported sleep logs, as non-invasive techniques, have been widely used to monitor human sleep-wake patterns. However, neither approach is a gold standard, and each has drawbacks. For example, actigraphy may overestimate sleep time and sleep efficiency resulting in substantial misclassification of the sleep state (e.g., individuals may have low activity over short time periods while fully awake) [18]. Sleep-logs are subject to inaccuracy and incompleteness, particularly in teenagers. This paper develops a hidden Markov modeling approach that uses both measures to estimate the sleep-wake cycle in teenagers.

Given that we expect activity measurements and self-reported sleep to relate to a person's underlying sleep state, we explore hidden Markov models (HMM) to characterize the sleep-wake cycle. HMM have been implemented in a variety of applications such as analyses of DNA sequences [17], epileptic seizure counts [2], speech recognition [28] and fetal lamb movement counts [22]. In addition, HMM have been used to model relapsing-remitting multiple sclerosis lesion counts in the brain [3, 4]. More recently, HMM have been used in treatment of alcoholism where the underlying true drinking status alternates between drinking and abstinence states [29]. A multi-profile two-level hidden Markov model was examined in a longitudinal childhood obesity study in which the model considers several states ("superstates") that characterize individual's psychosocial, food intake behavior, and physical activity behavior [16]. Despite its wide use in many areas, few studies were found that estimate the sleep-wake cycle using HMM. For example, [13] used electroencephalograms (EEG) to characterize the sleep-wake cycle and dealt with estimating sleep fragmentation using a HMM.

For longitudinal HMM, Bartolucci, F. & Solis-Trapala, I. [8] can be leveraged to illustrate the use of a multidimensional extension of latent Markov models to analyze data from studies with repeated binary responses in developmental psychology. Bartolucci, F. et al. [6] demonstrate the formulation of latent Markov models and the practical use of these models. The methods developed in Bartolucci, F. et al. [6] is well documented in the R package 'LMest' [7]. In addition, a comparison of some criteria for states selection in latent Markov models for longitudinal data can be found in Bacci S. et al. [5].

In the current work, we attempted to re-construct the sleep-wake cycle using HMM. The true unknown sleep state is not observed (hidden or latent) and we observe manifestations of sleep via two measures: actigraphy and sleep-logs. In this work, we extend the basic HMM to allow for bivariate emissions or outcome distributions as well as introduce non-homogeneous transition probabilities. We briefly introduce the data in Section 2, formulate the model and discuss parameter estimation in Sections 3 and 4, respectively. We fit the hidden Markov model for a series of participants in Section 5, perform simulation studies examining the proposed methodology in Section 6, and finally discuss the results and draw conclusions in Section 7.

## 2. Data

The participants were a subsample (called NEXT Plus) of overweight or obese (N = 282) and normal-weight (N = 278) adolescents in the NEXT Generation study (total N = 2,734), a seven-year longitudinal, nationally-representative study starting with 10th grade students in the 2009-2010 school year. Parental consent and participant assent were obtained. The study protocol was reviewed and approved by the Institutional Review Board of the *Eunice Kennedy Shriver* National Institute of Child Health and Human Development.

Two assessments were conducted to monitor adolescents' sleep patterns. Adolescents were asked to wear an actigraph (Actiwatch2 activity monitor) on their non-dominant wrist for 24 hours/day for seven consecutive days to record activity counts in 30 sec epochs, which were converted to 1-minute epochs to provide 1440 observations per day. Over the same time period, students were asked to indicate their sleep and wake times recorded to the minute. Self-reported observations were categorized into 1-minute intervals for 24 hours in a day. That is, each minute of actigraph and self-report data was quantified as either sleep (1) or wake (0). In total there were a possible 10080 actigraphy and self-report observations for each student. Proportion of self-report with missingness is 0.26. That is, about 26% of the self-report observations across all subjects is missing. There were no missing observations for activity.

In addition to missing self-reported sleep data, both actigraphy and self-reported sleep data may each be subjected to measurement error or misclassification. A combination of both actigraphy and self-report is more likely to result in a better estimate of sleep parameters that deal with both the missing self-reported data as well as the measurement/misclassification error [18]. Of interest to NEXT investigators is to obtain an unbiased estimate of the sleep-wake cycle for each teenager in the study. This assessment is expected in future studies to better examine effects of the sleep-wake cycle on subsequent clinical and behavioral outcomes.

Two subjects' data were selected for the analyses in this study. Subject 1 has complete self-report data while subject 2 displays missing self-report assessments.

## 3. Model

The hidden Markov model is formulated on an individual by individual basis. For each individual, define $X_t$ and $Y_t$, $\{X_t, Y_t: t = 1, \ldots, T\}$, as self-reported sleep assessment and the activity counts, respectively. Define $C_t$ as the underlying sleep state ($C_t$ =1 and 0, corresponding to sleep and awake, respectively) which is assumed to follow a Markov chain. The observed data are modeled as independent conditional on the sequence of hidden states. In addition, no cross-dependence is assumed between the observed data, $X_t$ and $Y_t$, conditional on the sequence of hidden states. Specifically, if the observed data $\{(X_t, Y_t): t = 1, \ldots, T\}$ follows a HMM then

i) The hidden states, $\{C_t: t = 1, \ldots, T\}$, follow a Markov chain.

ii) $\{(X_t, Y_t): t = 1, \ldots, T\}$ are conditionally independent given the hidden or latent states. That is, given $C_t$, $(X_t, Y_t)$ are independent of $X_1, \ldots, X_{t-1}, X_{t+1}, \ldots, X_T$ and $Y_1, \ldots, Y_{t-1}, Y_{t+1}, \ldots, Y_T$.

To summarize the relationship, given the current state of $C_t$, the observed variables, defined $X_t$ and $Y_t$, are conditionally independent from previous and future observations and states. Briefly, the Markov property is $P(C_{t+1}|C_t, \ldots, C_1) = P(C_{t+1}|C_t)$ which says that the conditional probability distribution of future states (the state at time $t+1$) depends only on the present state (the state at time $t$) and not on the preceding states. Further, conditional independence assumption postulates that the observed data is independent across time given the hidden state. Measurements were taken at each minute $t$ for 7 days for each person.

The full likelihood for T successive observations is constructed from 2 sets of random variables with realizations $x_1, x_2, \ldots, x_T$, and $y_1, y_2, \ldots, y_T$ which are assumed to be generated by a 2-state HMM. The random variables $X_t$ and $Y_t$ correspond to self-reported assessments and activity counts, respectively where $X_t$ is assumed to follow a binary distribution and $Y_t$ is assumed to follow a negative binomial. A negative binomial, rather than a Poisson distribution, is chosen for the count data since there is apparent over-dispersion in the activity count data. The joint distribution of the random variables is given by:

$$L_T = P(\{X_t, Y_t: t = 1,2, \ldots T\}) = \sum_{C_1, C_2, \ldots, C_T} P(X_1, X_2, \ldots, X_T, Y_1, Y_2, \ldots, Y_T, C_1, C_2, \ldots, C_T). \quad (1)$$

The components required for the construction of HMM likelihood are as follows:
i) Assume initial state probabilities $Pr(C_1 = i) = \delta_i$ for $i = 0, 1$. So, $Pr(C_1 = 0) = \delta_0$ and $\delta_1 = 1 - \delta_0$.
ii) Define the transition probabilities as $\gamma_{ij}(t) = Pr(C_{t+1} = j|C_t = i)$ which is modeled on the logit scale as $\text{logit}\gamma_{00}(t) = \alpha_0 + \alpha_1 \cos\left(\frac{2\pi t}{1440}\right) + \alpha_2 \sin\left(\frac{2\pi t}{1440}\right)$ and $\text{logit}\gamma_{11}(t) = \beta_0 + \beta_1 \cos\left(\frac{2\pi t}{1440}\right) + \beta_2 \sin\left(\frac{2\pi t}{1440}\right)$. The logit transformation is chosen for the transition probabilities since it constrains the probabilities so they are between 0 and 1. Since the sleep-wake cycle is clearly circadian, we model the transition probabilities (on the logit scale) with a sinusoidal curve over a one-day period. We assume that, conditional on the unknown sleep state, the activity count $Y_t$ follows a negative binomial distribution and the measured self-reported sleep indicator $X_t$ follows a Bernoulli distribution. Specifically, $Y_t = y_t | C_t = i \sim NegBin(s_i, \mu_i)$, $X_t = x_t | C_t = i \sim Bin(1, \pi_i)$ and $x_t$, $y_t$ are realizations at each time point. The negative binomial formulation allows for overdispersion relative to a Poisson distribution where $s_i$ and $\mu_i$ are the size and mean parameters. The relationship between these parameters and the variance of the activity counts is described via the formula $\mu_i \left(1 + \frac{\mu_i}{s_i}\right)$. The ratio, $\frac{\mu_i}{s_i}$, characterizes overdispersion. Further, $P(X_t = 1 | C_t = 0) = \pi_0$ is one minus the specificity and $P(X_t = 1 | C_t = 1) = \pi_1$ is the sensitivity of the self-rating for detecting the sleep state.

The complete-data likelihood can be written as:

$$P(X^{(T)}, Y^{(T)}, C^{(T)}) = P(C_1)P(X_1|C_1)P(Y_1|C_1) \times$$
$$\prod_{t=2}^{T}\{P(C_t|C^{(t-1)}, X^{(t-1)}, Y^{(t-1)})P(X_t|C^{(t)}, X^{(t-1)}, Y^{(t-1)})P(Y_t|C^{(t)}, X^{(t-1)}, Y^{(t-1)})\}, \quad (2)$$

where $X^{(t)}$ and $Y^{(t)}$ denote history up to time $t$ random variables.

By Markov property, (2) can be further simplified to yield,
$P(C_1)P(X_1|C_1)P(Y_1|C_1) \prod_{t=2}^{T}\{P(C_t|\gamma_{00}, \gamma_{01}, \gamma_{10}, \gamma_{11}, C_{t-1})P(X_t|\pi_0, \pi_1, C_t)P(Y_t|\mu_0, \mu_1, s_0, s_1, C_t)\}$. Let $\delta_{c_1} = P(C_1)$, $p_{c_1}(x_1) = P(X_1|C_1)$, and $p_{c_1}(y_1) = P(Y_1|C_1)$. Then we can rewrite the likelihood function as follows:

$L_T = \sum_{c_1,\ldots,c_T}\left[\delta_{c_1}p_{c_1}(x_1)p_{c_1}(y_1)\prod_{t=2}^{T}\{\gamma_{c_{t-1},c_t}(t)p_{c_t}(x_t)p_{c_t}(y_t)\}\right]$. Using the notation of Zucchini and MacDonald [35], define the forward probability $a_1(j) = \delta_j p_j(x_1)p_j(y_1)$ for $t = 1$ and $\gamma_{ij}(t)p_j(x_t)p_j(y_t)$ for $t = 2, 3, \ldots, T$. Thus, only the forward probability is involved in the evaluation of the likelihood.

## 4. Parameter Estimation

Our model is a two-state HMM fitted by direct numerical maximization (DNM) of the likelihood. The model parameters that we are interested in estimating are: $\alpha_0, \alpha_1, \alpha_2, \beta_0, \beta_1, \beta_2, \pi_0, \pi_1, s_0, s_1, \mu_0$, and $\mu_1$.

To perform maximum likelihood estimation in HMMs, two popular methods are the expectation-maximization (EM) and direct numerical maximization (DNM). We resort to DNM as it produces the MLEs faster (CPU time) than the EM algorithm [4]. In DNM, the R subroutine "nlm" is used to carry out maximization of the log-likelihood via Newton-type algorithm. We re-parameterized the initial parameters $\pi_0$ and $\pi_1$ on the logit scale and the mean actigraphy counts $\mu_0$ and $\mu_1$ on the log-scale, recognizing necessary constraints on these parameters. Likewise, the transition probabilities are conveniently re-parameterized using the logit transformation.

Estimation of the posterior probabilities (probability of true sleep at different times during the day) are carried out via recursive evaluation of the forward-backward probabilities [9]. Specifically,

$$E\left(C_t | X^{(T)}, Y^{(T)}\right) = \frac{P(X_t, Y_t, C_t=1)}{\sum_{l=0}^{1} P(X_t, Y_t, C_t=l)} = \frac{a_1(t) b_1(t)}{\sum_{l=0}^{1} a_l(t) b_l(t)}, \quad (3)$$

where $a_l(t)$ and $b_l(t)$ are arrays of forward and backward probabilities defined as:

$$a_l(t) = \begin{bmatrix} a_0(1) & a_1(1) \\ a_0(2) & a_1(2) \\ \vdots & \vdots \\ a_0(n) & a_1(n) \end{bmatrix}, \quad b_l(t) = \begin{bmatrix} b_0(1) & b_1(1) \\ b_0(2) & b_1(2) \\ \vdots & \vdots \\ b_0(n) & b_1(n) \end{bmatrix}.$$

Succinctly, $a_l(t) = \sum_{l=0}^{1} P(X_1^t, Y_1^t, C_{(t)} = l) = \sum_{l=0}^{1} \sum_{i=0}^{1} a_i(t) \gamma_{li} P(X_1^t | C_t = l) P(Y_1^t | C_t = l)$ and $b_l(t) = \sum_{l=0}^{1} P(X_{t+1}^n, Y_{t+1}^n, C_t = l) = \sum_{l=0}^{1} \sum_{i=0}^{1} b_i(t) \gamma_{li} P(X_{t+1}^n | C_t = l) P(Y_{t+1}^n | C_t = l)$ where $X_1^t = (X_1, X_2, \ldots, X_t)$ and $Y_1^t = (Y_1, Y_2, \ldots, Y_t)$.

The expression (3) evaluates the conditional expectation of the sleep/wake status at time t given the two sequences of observed data. This has been referred to as local decoding since it is computed separately for each *t* without explicit consideration of the sleep/wake cycle at other time points. Global decoding which can be evaluated using the Viterbi algorithm [31] chooses the sequence of hidden states that is most likely given the data. For the case of two states, these two different types of decoding should result in similar performance.

In the results section we fit the probability of true sleep and show reconstruction of sleep-wake cycle for two cases from our dataset.

The key interest is estimating the probability of sleep at a minute by minute basis for each individual $P(C_t = 1 | X_t, Y_t)$ for $t = 1, \ldots, T$. As such we estimate the probability of sleep for two subjects in our dataset as well as some simulated examples.

Since the advent of HMM by Baum et. al. [9], the EM algorithm has been the choice of estimation method to optimize the maximum likelihood function and fit the model. In recent years, both EM and DNM algorithms are two popular estimation methods in HMM as described in the book by Zucchini and MacDonald [35]. But availability of optimization routines in R software and computational ease make direct maximization a practical choice. MacDonald [25, 26] presented several examples to show the advantages of DNM over EM. He contends that DNM can be a viable alternative to EM algorithm and even argued for the utilization of DNM as its implementation requires minimal coding efforts. Bulla and Berzel [11] are among the early advocates of DNM method. Turner [30] also provides examples to show the ease, speed and straightforward application of the DNM (using Levenberg–Marquardt algorithm) as compared to the EM algorithm. In addition, Altman and Petkau [4] discussed the use of DNM in the practical setting of modeling brain lesions in relapsing-remitting multiple sclerosis.

Starting values were obtained by estimating parameters associated with the emission distribution for the over-dispersed activity and self-reported binary data by estimating these values over time periods from 1am to 6am for sleep and 1pm to 6pm for awake. For the latent process, various starting values were tried across the parameter space.

INSERT TABLE 1 HERE.

INSERT TABLE 2 HERE.

## 5. Results

In this section, we fit a HMM separately for each individual and subsequently, illustrate the methodology with two individuals from our dataset (Figures 1 and 2, respectively). Subject 1 has complete self-report data, while subject 2 displays missed assessments in the self-report. The parameter estimates for both subjects suggested a high diagnostic accuracy of self-reported sleep (the self-rating parameter estimates, $\pi_0$ and $\pi_1$, detected the sleep states accurately). Observed Hessian matrix was produced from the 'nlm' optimization procedure in R. Accordingly, the standard error of parameter estimates for both subjects were included. The HMM derived from activity counts and self-reported sleep seems to reconstruct the sleep-wake cycle realistically.

INSERT FIGURE 1 HERE.

INSERT FIGURE 2 HERE.

The mean count for awake and sleep are denoted as $\mu_0$ and $\mu_1$. The variance of the count formula, $\mu\left(1 + \frac{\mu}{s}\right)$, makes clear that a large value of $s_0$ corresponds to small amounts of overdispersion (relative to a Poisson model) for the individual in the awake state. Thus, the count distribution is Poisson when s gets very large (infinity). As mentioned previously, a value of $\pi_{i0}$ close to zero is consistent with the first count (or series of counts) being small and suggesting the very high likelihood that the individual is sleeping.

The observation period starts at 12am, so it is likely that an individual is sleeping at this time. The fact that the estimates for $\pi_{i0}$ and $\pi_{i1}$ are close to the boundaries simply imply strong evidence of that individual sleeping at midnight. The corresponding small variance also makes sense from the property of a binary random variable. When a binary random variable is very close to 1 or 0, it will have small variability. For example, the variation for a binomial proportion when it is 0.999 will be close to zero.

## 6. Simulation

In this section, we conduct simulations to show that we can accurately detect features of the sleep-wake cycle with the HMM. Evaluating the technique, we simulate the true sleep wake cycle as a fixed binary process that mimics regular sleeping with and without naps.

We simulate under the following fixed sleep-wake cycles:

1- Regular sleeping: wake up at 8 AM and go to sleep at 12 AM all seven days.

2- Regular sleeping: wake up at 6 AM and go to sleep at 12 AM all seven days with naps from 4 PM to 6 PM on all days.

3- Same as 2, but naps from 4 PM to 5 PM on all days.

4- Same as 2, but naps from 4 PM to 4:30 PM on all days.

5- Same as 2, but nap from 4 PM to 6 PM on only two of the seven days.

Although the model assumes that the sleep-wake cycle follows a Markov chain, we simulated according to a fixed cycle for several reasons. First, the pattern is fixed across all the simulated realizations making it possible to evaluate the performance of the method for particular sleep/wake configurations. Second, some realizations from a Markov chain may not result in sensible sleep/wake patterns. For this reason, the parameters that characterize the Markov chain are treated as nuisance parameters and are not themselves evaluated in the simulation.

We considered sensitivity and specificity of self-reports of 0.95 and 0.95. For the negative binomial distribution, we use parameters of one of the two individuals. We repeated all simulations on complete data and with ratings being missed on the last 3 days. We simulate 500 realizations for each case. Figures 3 and 4 display simulations with complete and missing ratings, respectively. Each figure also shows the fixed pattern, the mean posterior probability and the difference between the upper and lower 95% confidence interval for the mean posterior probability. As can be noted, the HMM reconstructs the sleep cycle.

INSERT FIGURE 3 HERE.

INSERT FIGURE 4 HERE.

The plot with the difference in the $2.5^{th}$ and $97.5^{th}$ percentile as a function of time for each scenario gives an assessment of the uncertainty around the estimated mean posterior probability. For scenario 1 in Figures 3 and 4 (data not shown), the maximum difference was 0.022 suggesting that the methodology computes the posterior means with very high accuracy. However, for some scenarios this interval was wider. For scenarios 2-4 in Figure 4 (data not shown), where we generated data with repeated naps, the difference in the limits was as high as 0.052 still suggesting relatively high precision in these posterior estimates.

The simulations suggest that we are able to reconstruct the sleep pattern with very high accuracy with our approach. This included estimating naps as well as when there is sizable amount of missing data in the self-reports. The length of the 95% confidence intervals is very narrow when the self-ratings are measured (all of Figure 3 and first four days in Figure 4). Interestingly, although the difference corresponding to the last three days where self-ratings are not measured (Figure 4) are larger than during the first four days, they are still quite narrow. This suggests that designs where self-reports are only partially observed over follow-up may be a cost-effective alternative to observing intensive self-reporting over complete follow-up. Further, the small amounts of error in these posterior probabilities and the fact that they are very close to zero or one also suggests that we can simply use plug-in regression calibration using these posterior probabilities to replace the true probabilities in subsequent NEXT analyses when relating sleeping patterns to poor behavioral outcomes.

The simulation presented in this paper evaluates how well the approach can identify particular sleep-wake patterns. By fixing these patterns, we guarantee the same pattern for each simulated realization, making an assessment of posterior probability and onset time identification transparent. If we generated the patterns differently for each realization (i.e., we generated according to a Markov chain), the evaluation would be cumbersome to do (e.g. naps may not occur in particular realizations of a Markov chain). The strength of our simulation approach is identification of state-occupancy (sleep/awake) and precise estimation of onset times even when the Markov structure is not correct.

## 7. Discussion and Conclusion

Obesity risks, including decreased daytime activity, increased appetite and pathological changes similar to the metabolic syndrome are also associated with short sleep duration and untreated sleep disordered breathing [19]. Decreased total sleep time among adolescents is exceedingly common. Research is needed to elucidate causal mechanisms linking physical activity, sedentary behavior, diet and sleep disturbance to the risk of overweight and CVD in adolescents and to investigate potential opportunities to reduce these risks. Combining actigraphy with self-report appears to substantially improve the accuracy of estimates of parameters of sleep [35]. We developed a new hidden Markov modeling approach for estimating sleep-wake cycle from actigraphy and self-report data. The model allowed for missing at random (MAR) [23] data in self-report.

Combining multiple information sources (i.e, actigraphy and sleep logs) using HMM is an innovative way to reconstruct sleep-wake cycle. The HMM can be used to reconstruct the sleep-wake cycle. They can also account for measurement error and deal with missing data while treating sleep as a covariate in longitudinal analyses and incorporate additional data elements such as ambient light and use of accelerometers. Future work will focus on prospectively relating sleep patterns to health outcomes (cardiovascular risk factors, academic performance, job performance, depression and driving performance) [10, 10, 27]. Since the posterior probability estimates appeared to be very close to zero or one for all time points on all individuals analyzed with this method, simple plug-in calibration appears reasonable for this type of analysis. Alternatively, we can develop a joint model for estimating the relationship between sleep and the subsequent outcome if need be. This is an area of future research. With the HMM technique, we can also investigate whether self-report on all seven days is necessary for precise estimation of the sleep-wake cycle. For example, self-reporting a sample of days with actigraphy over seven days may be nearly as efficient as complete observation.

The scope of the analyses presented in the paper is limited to two states since the authors wanted to make explicit inference to the sleep-wake cycle. However, in a different context, one could use HMMs to look for underlying activity changes such as different types of activity. For such an application [15], one may want to fit models with a flexible number of states and use BIC to select the best fitting model.

The purpose of introducing a negative binomial distribution for the activity count distribution was to generalize the typical HMM model for counts that relies on a Poisson assumption. There has been little work on model diagnostics and model adequacy for HMMs. This is particularly true for bivariate emissions distributions such as in this article (activity and self-report sleeping). This is an area for future research.

A non-parametric model can be used to fit HMMs [1]. An extension of such an approach to a bivariate emissions distribution (where one of the distributions is a time series of counts) is an interesting area for future research.

We emphasize that similar to all finite mixture models, the HMM model is identifiable up to a label switching of the two latent states (can call sleep and wake states either by 1 and 0 or alternatively by 0 and 1). One empirical demonstration of model identification is in the fact that our approach is not very sensitive to starting values. Issues on the identification of HMMs is discussed in the literature [21].

**Acknowledgement**

This research was supported by the Intramural Research Program of the Eunice Kennedy Shriver National Institute of Child Health and Human Development (Contract # HHSN267200800009C), and the National Heart, Lung and Blood Institute (NHLBI), the National Institute on Alcohol Abuse and Alcoholism (NIAAA), and Maternal and Child Health Bureau (MCHB) of the Health Resources and Services Administration (HRSA), with supplemental support from the National Institute on Drug Abuse (NIDA).

24. H. G. Lund, B. D. Reider, A. B. Whiting & J. R. Prichard, Sleep patterns and predictors of disturbed sleep in a large population of college students, Journal of Adolescent Health 46 (2010), pp. 124-132.

25. I. L., MacDonald, Numerical Maximisation of Likelihood: A Neglected Alternative to EM?, International Statistical Review, 82 (2014): pp. 296-308.

26. I. L., MacDonald, Is EM really necessary here? Examples where it seems simpler not to use EM, AStA Adv Stat Anal 105 (2021), pp. 629–647.

27. L. R. McKnight-Eily, D. K. Eaton, R. Lowry J. B. Croft, L. Presley-Cantrell, G. S. Perry, Relationships between hours of sleep and health-risk behaviors in US adolescent students, Prev. Med 53 (2011), 271-3.

28. L. R. Rabiner, A tutorial on hidden Markov models and selected applications in speech recognition, Proceedings of IEEE 77 (1989), pp. 267–295.

29. K. E. Shirley, D. S. Small, K. G. Lynch, S. A. Maisto & D. W. Oslin, Hidden Markov Models For Alcoholism Treatment Trial Data. Ann. Appl. Stat, 4 (2010), pp. 366-395.

30. R. Turner, Direct maximization of the likelihood of a hidden Markov model, Computational Statistics & Data Analysis, 52 (2008), pp. 4147-4160.

31. A. J. Viterbi, Error bounds for convolutional codes and an asymptotically optimum decoding algorithm, IEEE Transactions on Information Theory, 13 (1967), pp. 260-269.

32. J. A. Williams, F. J. Zimmerman & J. F. Bell, Norms and trends of sleep time among US children and adolescents, Archives of Pediatrics and Adolescent Medicine 167 (2013), pp. 56-60.

33. L. J. Willson, J. L. Folks, & J. H. Young, Multistage Estimation Compared with Fixed-Sample-Size Estimation of the Negative Binomial Parameter k, Biometrics, 40 (1984), pp. 109-117.

34. K. Wulff, S. Gatti, J. G. Wettstein & R. G. Foster, Sleep and circadian rhythm disruption in psychiatric and neurodegenerative disease, Nature Reviews Neuroscience, 11 (2010), 589-599.

35. W. Zucchini & I. L MacDonald, Hidden Markov Models for Time Series: An Introduction Using R. CRC Press, Taylor & Francis Group, Boca Raton, 2009.

**Figure 1**: Subject 1 data and reconstructed sleep-wake cycle.

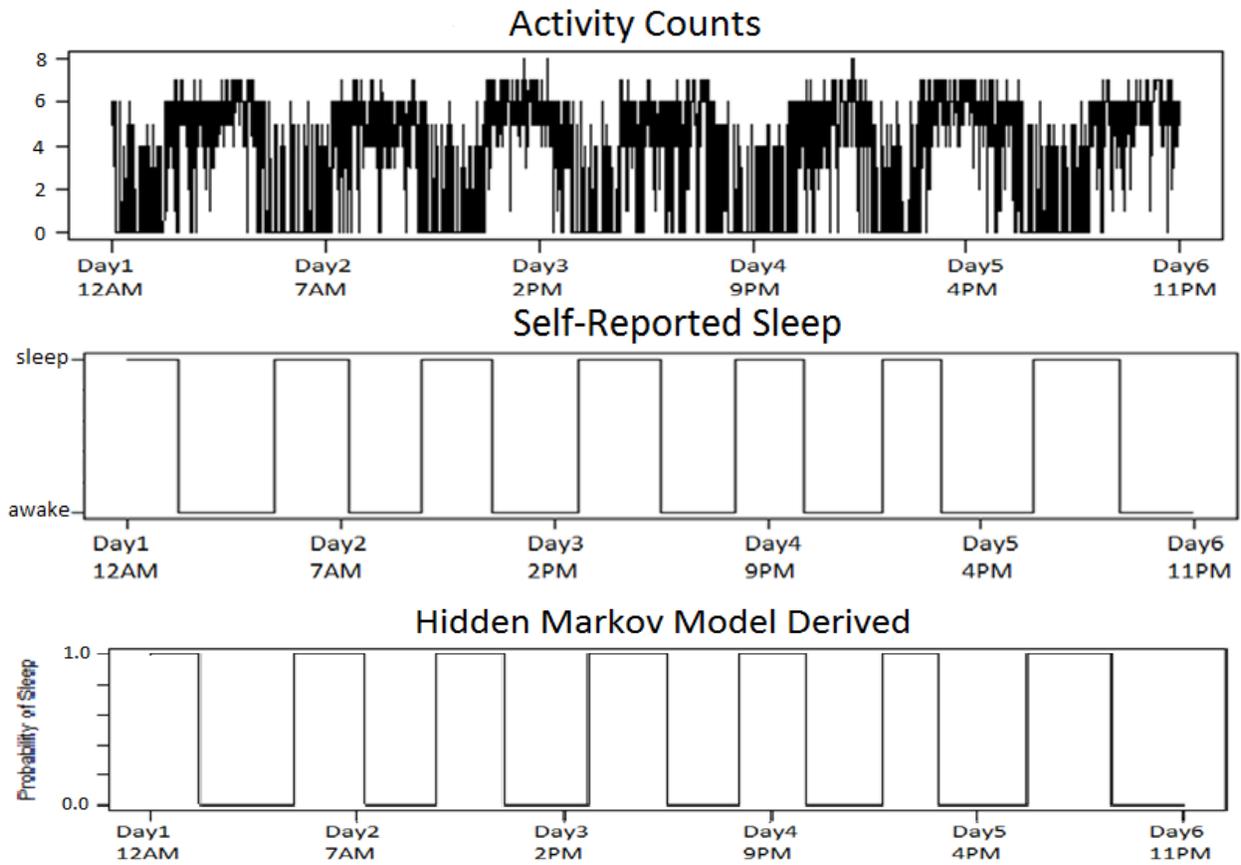

**Figure 2:** Subject 2 data and reconstructed sleep-wake cycle.

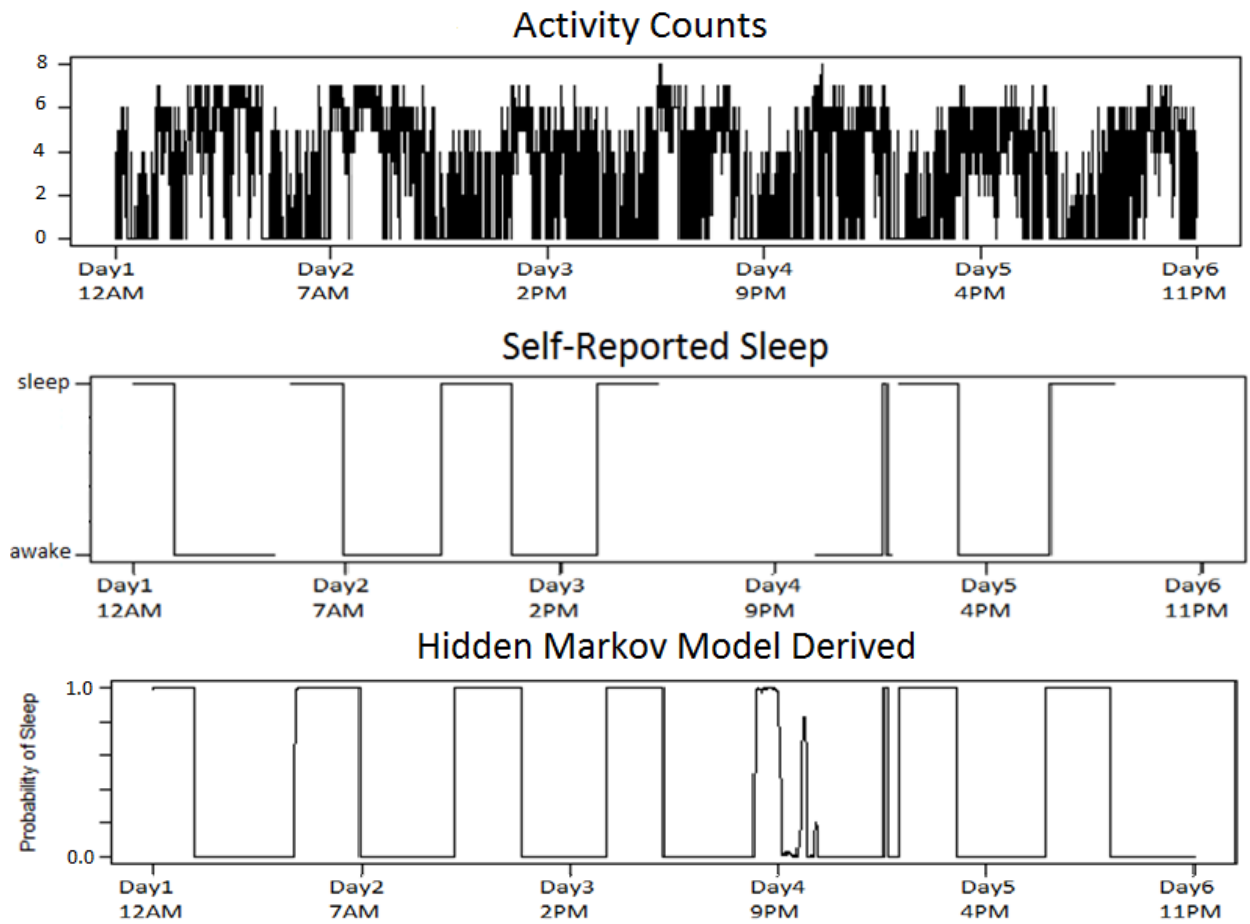

**Figure 3**: Complete fixed patterns 1-5.

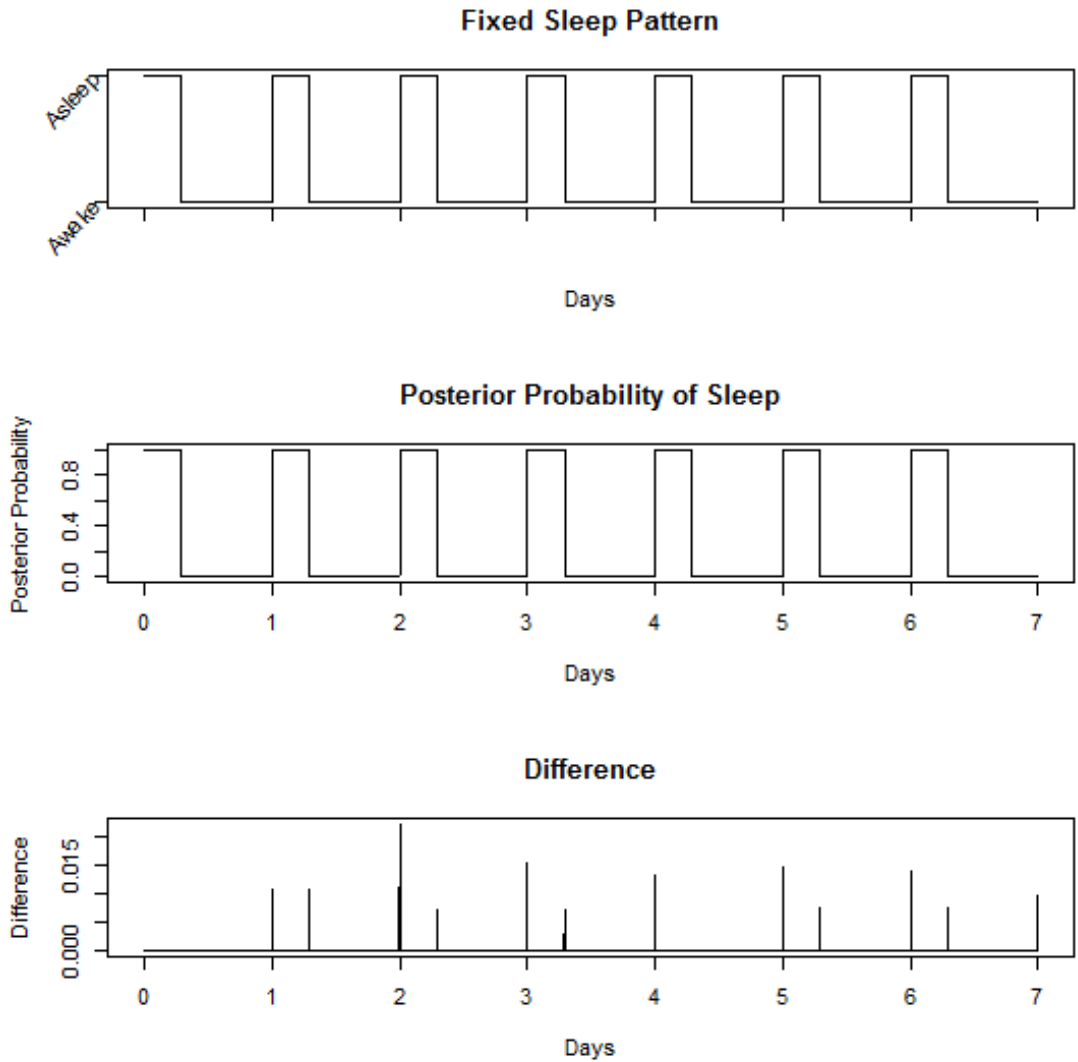

(1)

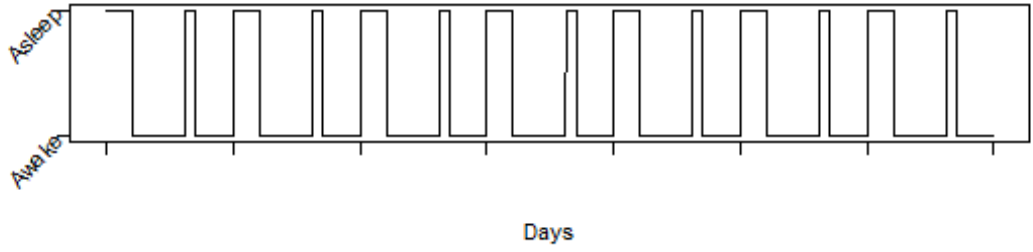

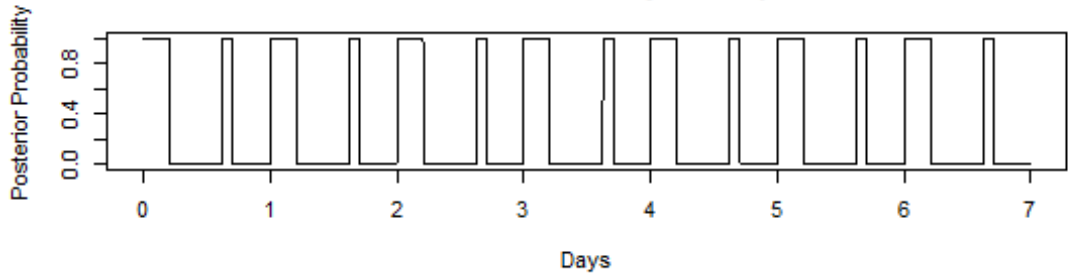

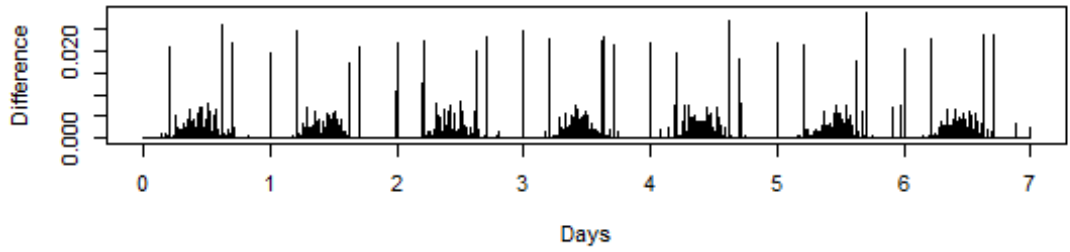

(2)

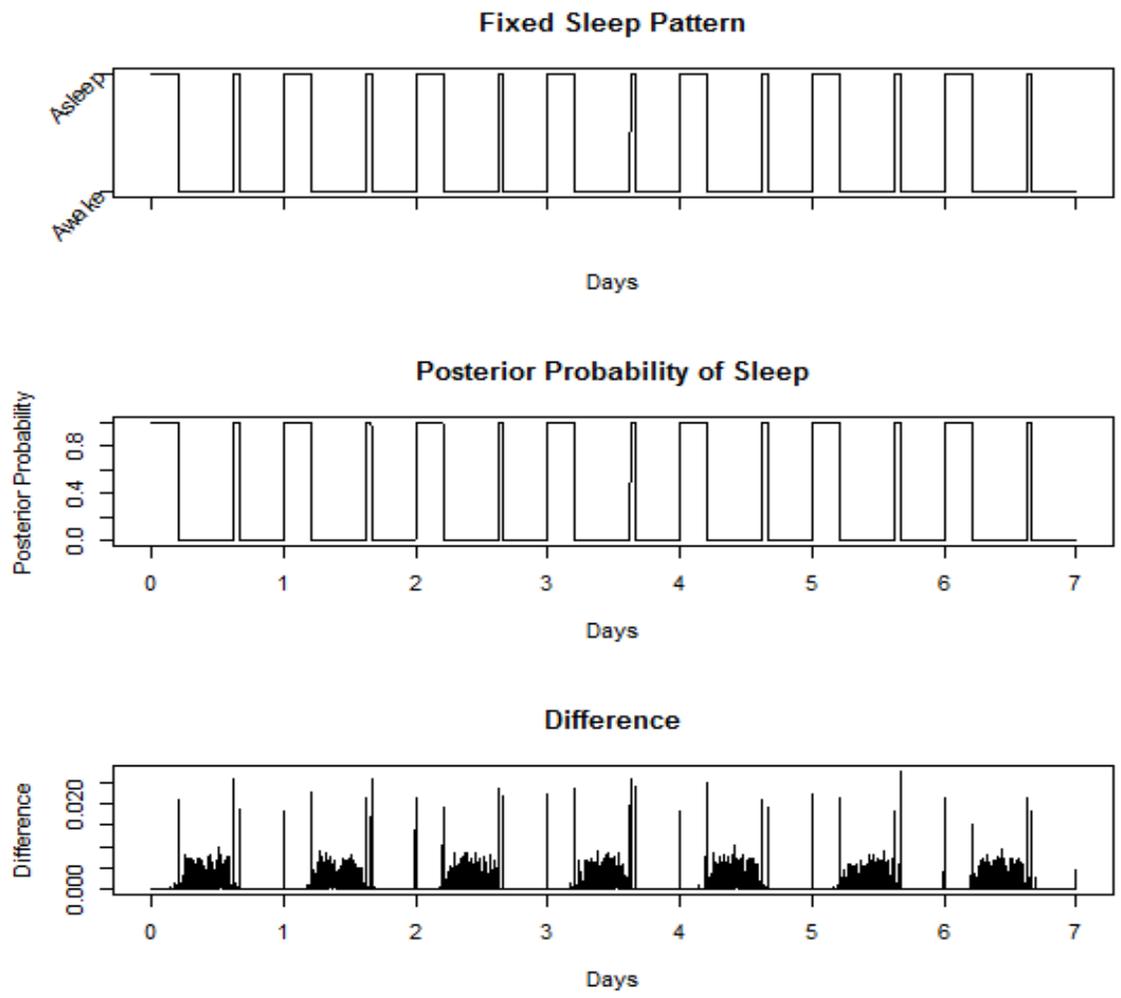

(3)

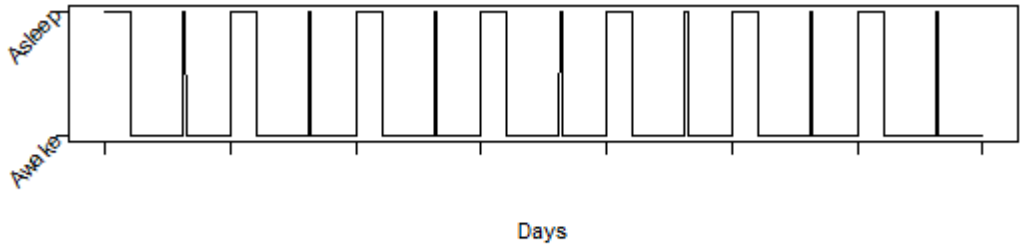

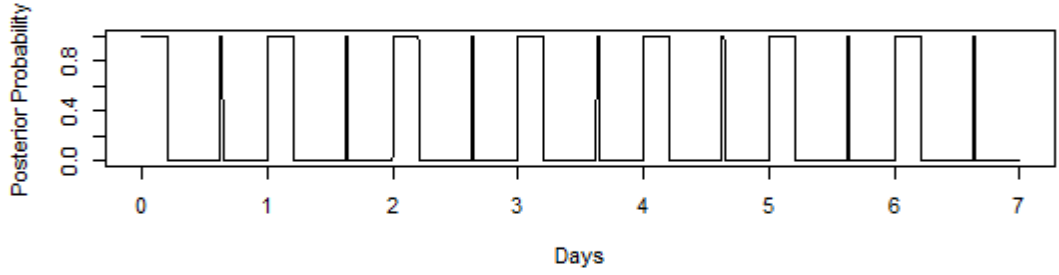

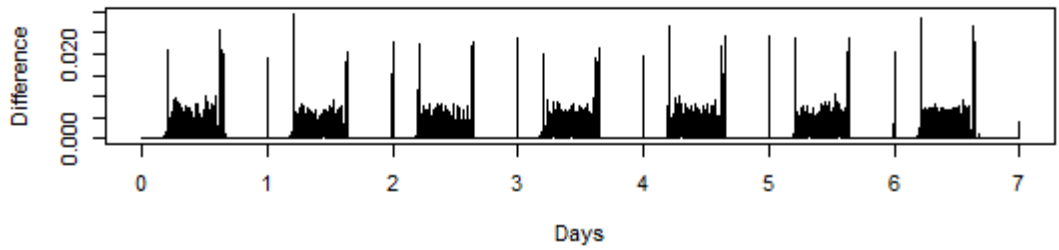

(4)

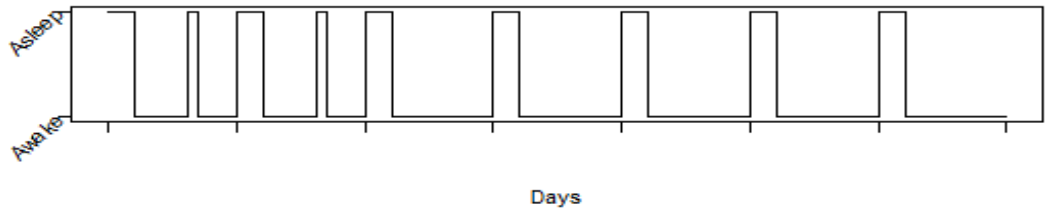

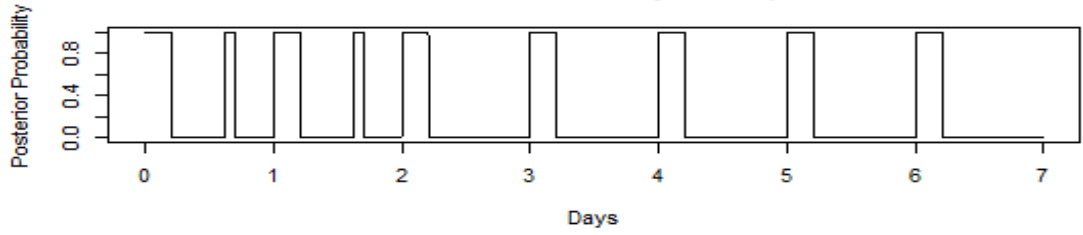

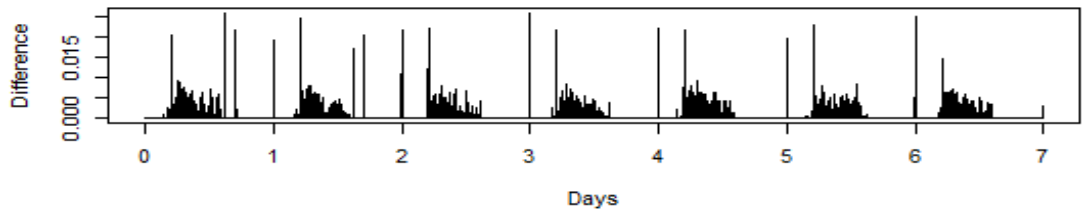

(5)

**Figure 4**: Missing fixed patterns 1-5.

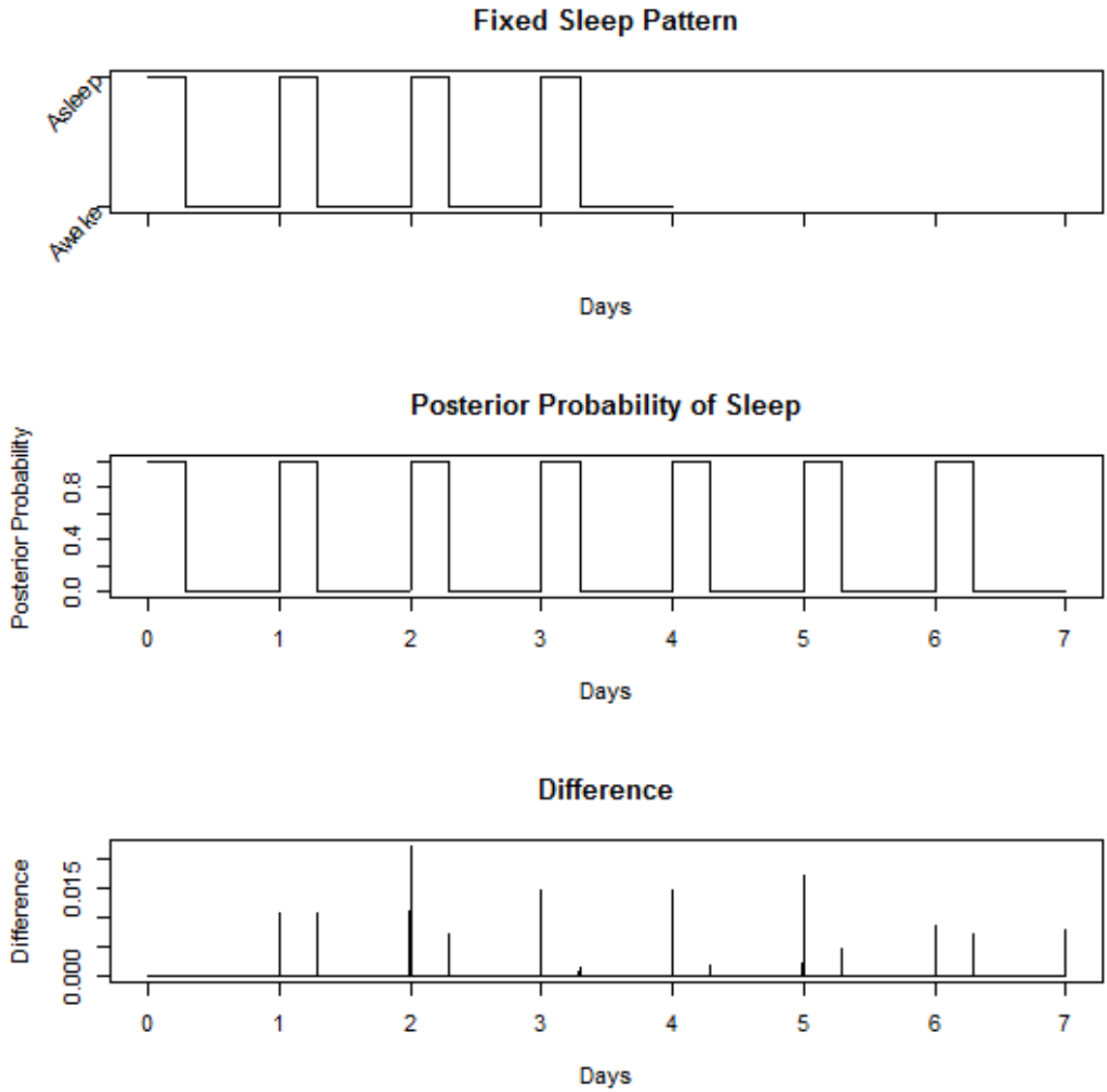

(1)

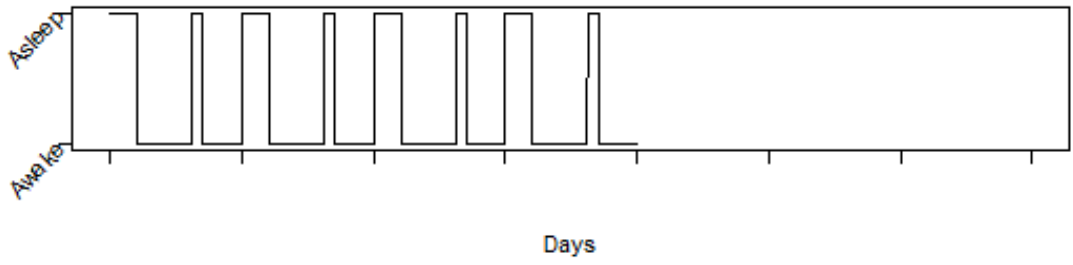
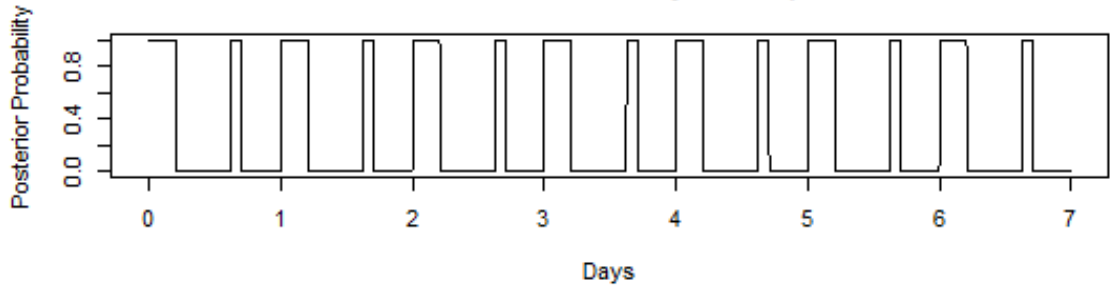
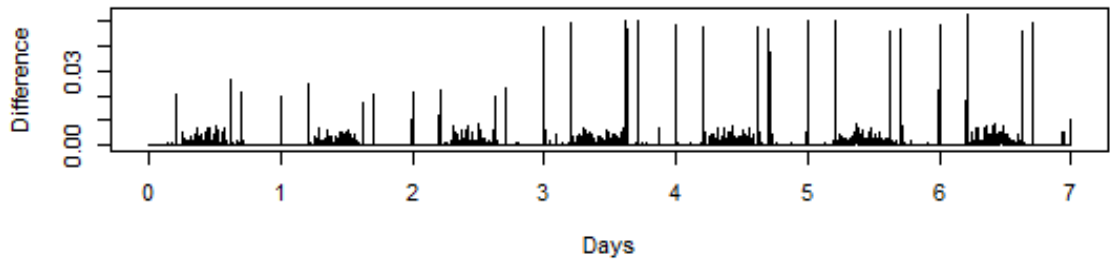

(2)

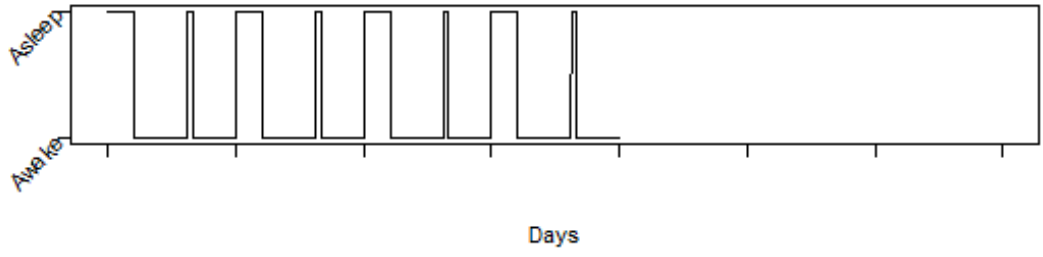

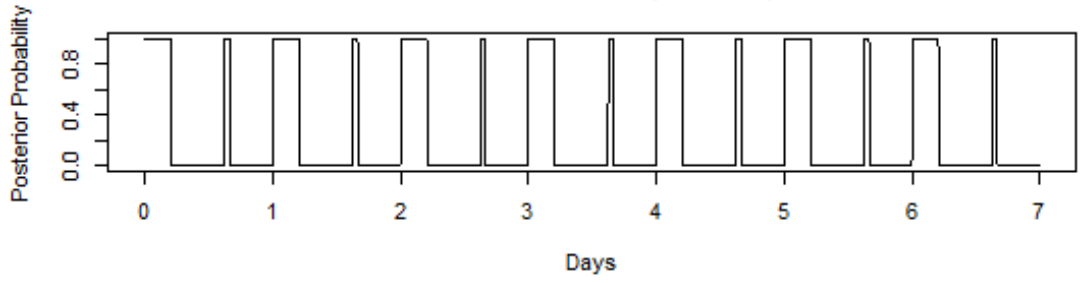

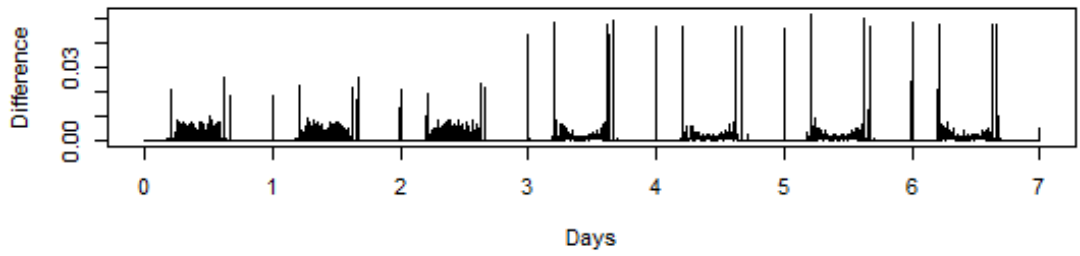

(3)

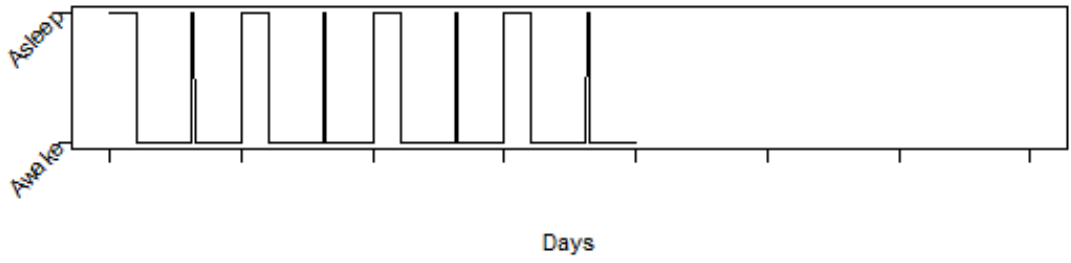

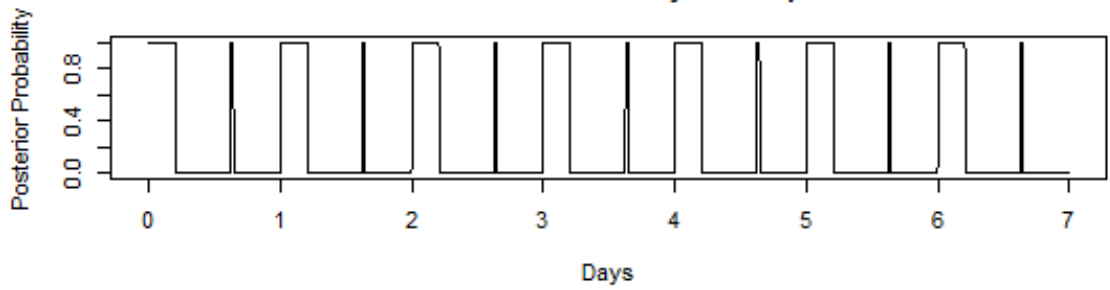

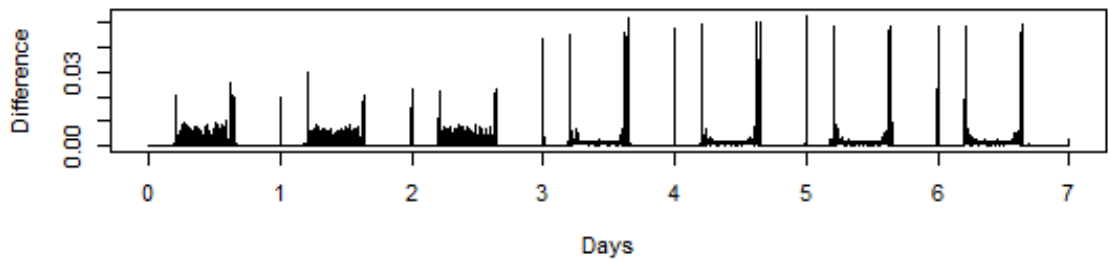

(4)

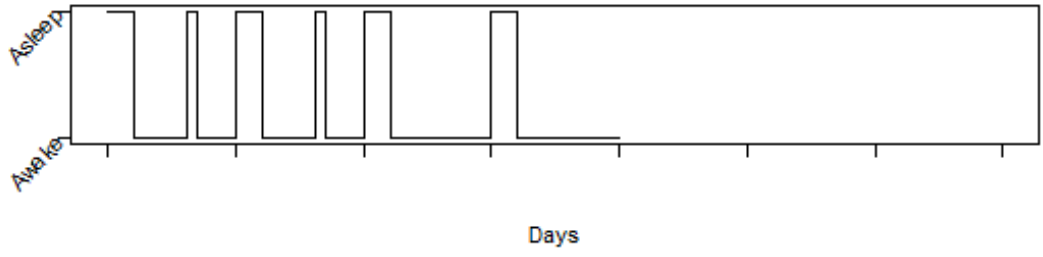

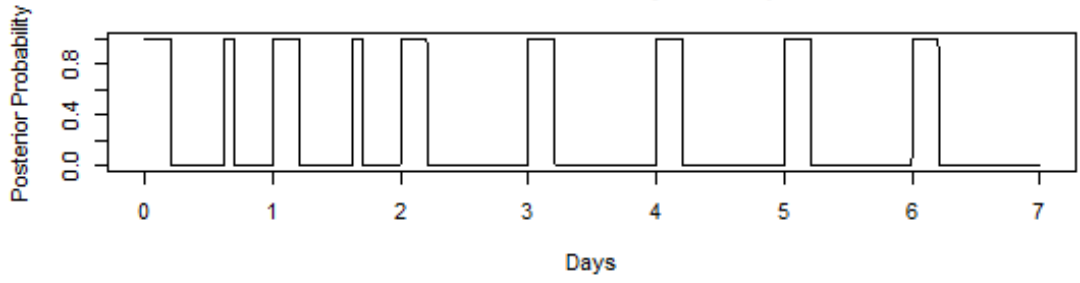

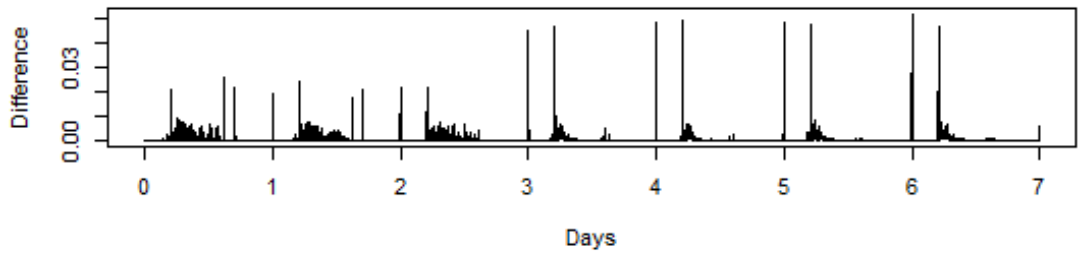

(5)

**Table 1**: Model fit for subject 1.

| Parameters | Estimates | SE* |
|---|---|---|
| $\mu_0$ | 5.19 | 0.0056 |
| $\mu_1$ | 1.31 | 0.067 |
| $s_0$ | 799 | 3.36 |
| $s_1$ | 0.18 | 0.064 |
| $\pi_0$ | 0.099 | 0.041 |
| $\pi_1$ | 0.99 | $5.54 \times 10^{-16}$ |
| $\alpha_0$ | 8.54 | 0.37 |
| $\alpha_1$ | -6.21 | 0.48 |
| $\alpha_2$ | -5.49 | 0.68 |
| $\beta_0$ | 7.31 | 0.35 |
| $\beta_1$ | 3.36 | 0.35 |
| $\beta_2$ | 1.06 | 0.82 |

**\* SE**=standard error of parameter estimates were computed using observed Hessian Matrix.

| Initial Values | Parameters | DNM Estimates |
|---|---|---|
| | $(\mu_0, \mu_1)$ | (5.19, 1.31) |
| $(\mu_0, \mu_1)$=(5, 10) | $(s_0, s_1)$ | (799, 0.18) |
| $(s_0, s_1)$=(4, 4) | $(\pi_0, \pi_1)$ | (0.099, 0.99) |
| $(\pi_0, \pi_1)$=(0.1, 0.99) | | |
| $\alpha_0 = 8$, $\alpha_1 = -5$, $\alpha_2 = -3$, | $(\alpha_0, \alpha_1, \alpha_2)$ | $(8.54, -6.21, -5.49)$ |
| $\beta_0 = 7.5$, $\beta_1 = 4$, $\beta_2 = 2$ | $(\beta_0, \beta_1, \beta_2)$ | $(7.31, 3.36, 1.06)$ |

Estimation of the shape parameter can be unstable occasionally due to small mean and large shape parameter. This phenomenon in negative binomial distribution has been investigated previously [33].

**Table 2:** Model fit for subject 2.

| Parameters | Estimates | SE* |
|---|---|---|
| $\mu_0$ | 4.58 | 0.00617 |
| $\mu_1$ | 0.82 | 0.043 |
| $s_0$ | $2.54 \times 10^8$ | 0.74 |
| $s_1$ | 0.15 | 0.047 |
| $\pi_0$ | $4.8 \times 10^{-8}$ | 7.99 |
| $\pi_1$ | 1 | 0.12 |
| $\alpha_0$ | 4.69 | 0.20 |
| $\alpha_1$ | -0.11 | 0.23 |
| $\alpha_2$ | -0.69 | 0.33 |
| $\beta_0$ | 3.81 | 0.26 |
| $\beta_1$ | 1.63 | 0.35 |
| $\beta_2$ | 3.17 | 0.31 |

**\* SE**=standard error of parameter estimates were computed using observed Hessian Matrix.

| Initial Values | Parameters | DNM Estimates |
|---|---|---|
| | $(\mu_0, \mu_1)$ | (4.58, 0.82) |
| $(\mu_0, \mu_1)$=(1.5, 0.2) | $(s_0, s_1)$ | $(2.54 \times 10^8, 0.15)$ |
| $(s_0, s_1)$=(20.17, 1.87) | $(\pi_0, \pi_1)$ | $(4.8 \times 10^{-8}, 1)$ |
| $(\pi_0, \pi_1)$=(0.0001, 0.8) | | |
| $\alpha_0 = 6$, $\alpha_1 = -0.1$, $\alpha_2 = -0.7$, | $(\alpha_0, \alpha_1, \alpha_2)$ | $(4.69, -0.11, -0.69)$ |
| $\beta_0 = 4$, $\beta_1 = 1.6$, $\beta_2 = 3$ | $(\beta_0, \beta_1, \beta_2)$ | $(3.81, 1.63, 3.17)$ |

Estimation of the shape parameter can be unstable occasionally due to small mean and large shape parameter. This phenomenon in negative binomial distribution has been investigated previously [33].